\documentclass[12pt]{iopart}

\usepackage{cite}  
\usepackage[colorlinks,linkcolor=blue,anchorcolor=blue,citecolor=blue,urlcolor=blue]{hyperref}
\usepackage{graphicx}
\usepackage{changes}
\usepackage{subfigure}
\usepackage{float}
\usepackage{threeparttable}
\usepackage{hyperref}
\usepackage{cleveref}
\begin{document}

\title{Systematic study of proton radioactivity based on Gamow--like model with a screened electrostatic barrier}

\author{Jiu-Long Chen$^1$,  Xiao-Hua Li$^{1,3,4}$ Jun-Hao Cheng$^1$, Jun-Gang Deng$^{1}$, and Xi-Jun Wu$^2$}

\address{$^1$School of Nuclear Science and Technology, University of South China, Hengyang 421001, China \\
 $^2$ School of Math and Physics, University of South China, Hengyang 421001, China\\
 $^3$ Key Laboratory of Low Dimensional Quantum Structures and Quantum Control, Hunan Normal University, Changsha 410081, China\\
 $^4$ Cooperative Innovation Center for Nuclear Fuel Cycle Technology $\&$ Equipment, University of South China, Hengyang 421001, China}

\ead{\mailto{lixiaohuaphysics@126.com}}
\ead{\mailto{wuxijun1980@yahoo.cn}}

\vspace{10pt}
\begin{indented}
\item[]
\end{indented}

\begin{abstract}
In the present work we systematically study the half--lives of proton radioactivity for $51 \leq Z \leq 83$ nuclei based on the Gamow--like model with a screened electrostatic barrier. In this model there are two parameters while considering the screened electrostatic effect of Coulomb potential with the Hulthen potential i.e. the effective nuclear radius parameter $r_0$ and the screening parameter $a$. The calculated results can well reproduce the experimental data. In addition, we extend this model to predict the proton radioactivity half--lives of 16 nuclei in the same region within a factor of 2.94, whose proton radioactivity are energetically allowed or observed but not yet quantified. Meanwhile, studying on the proton radioactivity half-life by a type of universal decay law has been done. The results indicate that the calculated half--lives are linearly dependent on Coulomb parameter with the same orbital angular momentum.
\end{abstract}

%
%
%
%
%

\section{Introduction}
\label{section 1}

The proton radioactivity was firstly observed in an isomeric state of $^{53}$Co in 1970 by Jackson et al. \cite{JACKSON1970281,CERNY1970284}. Subsequently, Hofmann et al. and Klepper et al. detected the proton emission from nuclear ground states in $^{151}$Lu \cite{Hofmann1982} and $^{147}$Tm \cite{Klepper1982} independently. With the development of experimental facilities and radioactive nuclear beams, the study of proton radioactivity is becoming one of hot topics in nuclear physics. Up to now, there are about 26 proton emitters decaying from their ground states and 15 different nuclei choosing to emit protons from their isomeric states, which have been identified between Z = 51 and Z = 83 \cite{1674-1137-41-3-030001,Qian2016}. As an important decay mode of unstable nuclei, proton radioactivity is an useful tool to obtain spectroscopic information because the decaying proton is the unpaired proton not filling its orbit and to extract important information about nuclear structure lying beyond the proton drip line, such as the coupling between bound and unbound nuclear states, the shell structure \cite{KARNY200852} and so on. There are a lot of models and empirical formulas having been proposed to deal with the proton radioactivity such as the effective interactions of density--dependent M3Y (DDM3Y) \cite{BHATTACHARYA2007263,QIAN-Yi-Bin-72301}, the single--folding model \cite{PhysRevC.72.051601,QIAN-Yi-Bin**-112301,QIAN-Yi-Bin-72301}, the distorted--wave Born approximation \cite{PhysRevC.56.1762},  Jeukenne, Lejeune and Mahaux (JLM) \cite{BHATTACHARYA2007263}, the generalized liquid--drop model \cite{PhysRevC.79.054330,0954-3899-37-8-085107,yzwang2017}, the finite-range effective interaction of Yukawa form \cite{Routray2011}, the R-matrix approach  \cite{PhysRevC.85.011303}, the Skyrme interactions \cite{Routray2012}, the relativistic density functional theory \cite{FERREIRA2011508},  the phenomenological unified fission model \cite{PhysRevC.71.014603,1674-1137-34-2-005}, the two--potential approach (TPA) \cite{Qian2016} which is also successfully applied to the $\alpha$ decay and cluster radioactivity \cite{PhysRevC.95.044303,PhysRevC.95.014319,PhysRevC.96.024318,1674-1137-41-12-124109,1674-1137-42-4-044102,doi:10.1142/S0218301318500052}, the Coulomb and proximity potential model (CPPM) \cite{PhysRevC.96.034619}, a simple empirical formula proposed by Delion et al. \cite{PhysRevLett.96.072501} and so on. For more details about different theories of proton radioactivity, the readers are referenced to Ref. \cite{delion424}.

Recently, Budaca et al. used a simple analytical model based on the WKB approximation considering the screened effect of emitted proton-daughter nucleus Coulomb interaction with the Hulthen potential to systematically study the half-lives of proton radioactivity for 41 nuclei with Z $\geq$ 51 \cite{Budaca2017}. The results indicate that the difference between the outer turning point radii corresponding to pure Coulomb and Hulthen barriers increases with the proton number $Z$ increasing. Whereas the penetration probability is sensitive to the outer turning point radii. In 2016, Zdeb et al. \cite{Zdeb2016} calculated the half-life of proton radioactivity with a Gamow--like model, which has much deeper physical basements and conserves the simplicity of the Viola--Seaborg approach \cite{PhysRevC.68.034319,PhysRevC.69.024614,PhysRevLett.96.072501,PhysRevC.78.044310,PhysRevC.79.054330}. In this work, considering the screened effect of emitted proton-daughter nucleus Coulomb interaction with the Hulthen potential, and combining with the phenomenological assault frequency, we modify the Gamow--like model proposed by Zdeb et al. \cite{Zdeb2016} and use this model to systematically study the half--lives of proton radioactivity for $51 \leq Z \leq 83$ nuclei. Meanwhile, we extend this model to predict the proton radioactivity half--lives of 16 nuclei in the same region, whose proton radioactivity are energetically allowed or observed but not yet quantified.

This article is organized as follows. In Section \ref{section 2}, the theoretical framework for the calculation of the proton radioactivity half--life is briefly described. The detailed calculations and discussion are presented in Section \ref{section 3}. Finally, a brief summary is given in Section \ref{section 4}.

\section{Theoretical framework}
\label{section 2}

The half--life of proton radioactivity is generically calculated by
\begin{equation}
T_{1/2}=\frac{\ln2}{\lambda}=\frac{\ln2}{\nu P}.
\end{equation}
The $\nu$ is the assault frequency related to the oscillation frequency $\omega$ \cite{PhysRevC.81.064309}. It can be expressed as
\begin{equation}
\nu=\frac{\omega}{2\pi}=\frac{(2n_r+{\ell}+\frac{3}{2})}{2\pi\mu R^2_n}=\frac{(G+\frac{3}{2})\hbar}{1.2\pi \mu R_0^2},
\end{equation}
where $R_n$ is the nucleus root-mean-square (rms) radius. The relationship $R_n^2=\frac{3}{5}R_0^2$ \cite{PhysRevC.62.044610} is used here. In this work, $R_0$ equates to the classical inner turning point defined as Eq. ({\ref{rin}}). $G=2n_r+\ell$ is the principal quantum number with $n_r$ and $\ell$ being the radial and angular momentum quantum number, respectively. For proton radioactivity we choose $G=$ 4 or 5 corresponding to the $4\hbar \omega$ or $5\hbar \omega$ oscillator shell depending on the individual proton emitter. $\mu=m_pm_d/(m_p+m_d)$ is the reduced mass of the decaying nuclear system with $m_p$ and $m_d$ being the mass of proton and daughter nucleus, respectively. $\hbar$ is the reduced Planck constant.

The penetration probability $P$ is calculated by the semi--classical Wentzel--Kramers--Brilloum (WKB) approximation same as the Gamow--like model \cite{Zdeb2016,zdeb014029,zedb024308} and expressed as
\begin{equation}
P=\exp\left[-\frac{2}{\hbar}\int_{R_{\rm{in}}}^{R_{\rm{out}}}\sqrt{2\mu(V(r)-E_p)}dr\right], 
\end{equation}
where $E_p$ is the kinetic energy of the emitted proton extracted from the released energy of proton radioactivity $Q_p$. It can be determined by the condition $E_p=Q_p(A_m-1)/A_m$ with $A_m$ being the mass number of the mother nucleus. $V(r)$ is the total emitted proton--daughter nucleus interaction potential. $R_{\rm{in}}$ is the classical inner turning point. $R_{\rm{out}}$ is the outer turning point from the potential barrier which is determined by the condition $V(R_{\rm{out}})=E_p$. In the Gamow--like model, $R_{\rm{in}}$ represents the radius of the spherical square well in which the proton is trapped before emission. It can be expressed as
\begin{equation}\label{rin}
R_{\rm{in}}=r_{0}A_d^{1/3}+R_{p}, 
\end{equation}
where $A_d$ is the mass number of the residual daughter nucleus. $R_{p}$ is the proton radius considered to be 0.8409 fm  in this work, whereas in Ref. \cite{Zdeb2016} it is factorized. $r_0$ is the effective nuclear radius constant, which is one of the adjustable parameters in this model. The range of $r_0$ is the same as the parameter in Ref. \cite{MYERS1974186,MYERS1991292}. 

\begin{figure}[!]
\centering
\includegraphics[width=8.5cm]{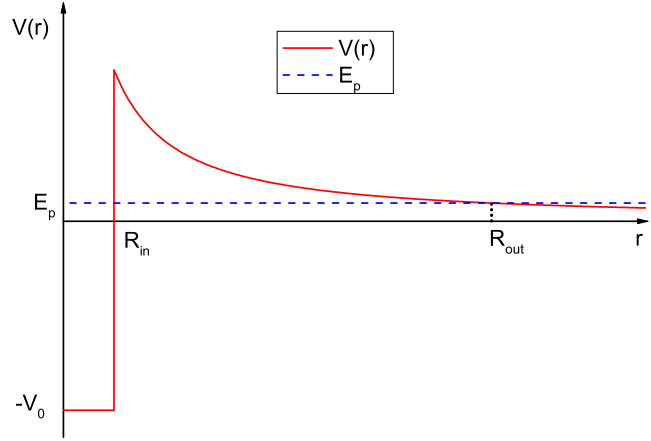}
\caption{\label{fig}(color online) Schematic plot of the potential energy $V(r)$ as a function of the distance between the nucleus and emitted proton in the Gamow--like model.}
\end{figure}

In general, the emitted proton-daughter nucleus electrostatic potential is by default of the Coulomb type $V_C(r)=Z_de^2/r$ with $Z_d$ being the proton number of daughter nucleus. Whereas, in the process of proton radioactivity, for the superposition of the involved charges, movement of the proton which generates a magnetic field and the inhomogeneous charge distribution of the nucleus, the emitted proton-daughter nucleus electrostatic potential behaves as a Coulomb potential at short distance and drop exponentially at large distance i.e. the screened electrostatic effect  \cite{Budaca2017}. This behaviour of electrostatic potential can be described as the Hulthen type potential which is widely used in nuclear, atomic, molecular and solid state physics\cite{Budaca2017,Hulthen1942,Hulthen1957,Oyewumi2016} and defined as 
\begin{equation}
V_H(r)=\frac{ae^2Z_d}{\exp(ar)-1}, 
\end{equation}
where $a$ is the screening parameter. In this framework, the total emitted proton--daughter nucleus interaction potential $V(r)$ shown in Fig. \ref{fig} is given by
\begin{equation}
\label{cases}
V(r)=\cases{-V_0&for $0 \leq r < R_{\rm{in}}$,\\
V_{H}(r)+V_{l}(r)&for $r \geq R_{\rm{in}}$,\\}
\end{equation}
where $V_0$ is the depth of the potential well. $V_{H}(r)$ and $V_{l}(r)$ are the Hulthen type of screened electrostatic Coulomb potential and centrifugal potential, respectively. 

Because $l(l+1)\to (l+\frac{1}{2})^2$ is a necessary correction for one--dimensional problems \cite{doi:10.1063/1.531270}, we adopt the Langer modified centrifugal barrier. It can be written as
\begin{equation}
V_l(r)=\frac{\hbar^2(l+\frac{1}{2})^2}{2 {\mu}r^2},
\end{equation}
where  $l$ is the orbital angular momentum taken away by the emitted proton. The minimum orbital angular momentum $l_{\rm{min}}$ taken away by the emitted proton can be obtained by the  parity and angular momentum conservation laws.

\section{Results and discussion}
\label{section 3}

\begin{figure}[!]
\centering
\includegraphics[width=8.5cm]{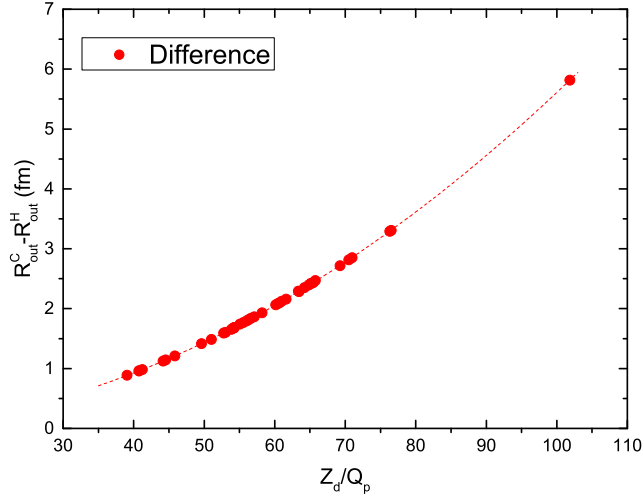}
\caption{\label{fig0}(color online) The difference between $R_{\rm{out}}^H$ and $R_{\rm{out}}^C$ which are obtained by the Gamow--like model where the electrostatic barrier are considered and unconsidered, respectively. The turning point radii are defined by $V^i(R_{\rm{out}})=E_p (i=C,H)$.}
\end{figure}

In the present work we systematically study the half--lives of proton radioactivity for the nuclei from $^{105}$Sb to $^{185}$Bi$^m$. The experimental proton radioactivity half--lives, spin and parity are taken from the latest evaluated nuclear properties table NUBASE2016 \cite{1674-1137-41-3-030001} except for $^{105}$Sb, $^{109}$I, $^{140}$Ho, $^{150}$Lu, $^{151}$Lu, $^{159}$Re$^m$, $^{164}$Ir, which are taken from Ref. \cite{Qian2016}, the proton radioactivity released energies are taken from the latest evaluated atomic mass table AME2016 \cite{1674-1137-41-3-030002,1674-1137-41-3-030003}. With the Hulthen potential for the electrostatic barrier being considered, our model contains two adjustable parameters i.e. the screening parameter $a$ and effective nuclear radius parameter $r_0$, which are obtained by fitting the experimental data of 41 proton emitters. The standard deviation $\sigma$ indicating the deviation between the experimental data and calculated ones can be expressed as
\begin{equation}
\sigma=\sqrt{\frac{1}{N}\sum\limits_{i=1}^N(\rm{log}_{10}T^{\it i}_{\rm{calc}}-\rm{log}_{10}T^{\it i}_{\rm{expt}})^2}.
\end{equation}
 According to the smallest standard deviation $\sigma_{\rm{cal1}}=0.468$, the two adjustable parameters are determined as $r_0=1.14$ fm and $a=5.6\times10^{-4}$ fm$^{-1}$, respectively. Their values are in the same range as in Ref. \cite{Budaca2017} and Ref. \cite{MYERS1974186,MYERS1991292,Zdeb2016}, respectively. Therefore, by considering the screened electrostatic effect of Coulomb potential with the Hulthen potential within a Gamow--like model, the calculated half--lives of the proton radioactivity in this work are within a factor of 2.94. For a more intuitive displaying the effect of non--zero electrostatic screening in the description of the proton emission phenomenon, we plot the variation of the difference between $R_{\rm{out}}^C$ and $R_{\rm{out}}^H$ with $Z_d/Q_p$ in Fig. \ref{fig0}. The outer turning points $R_{\rm{out}}^H$ and $R_{\rm{out}}^C$ are obtained by the Gamow--like model where the screened electrostatic effect are considered and unconsidered, respectively. The effect of non--zero screening in the proton radioactivity is sizeable as can be seen in Fig. \ref{fig0}. The screening of the electrostatic repulsion shortens this radius by several percents. Moreover, in case of Coulomb barrier, the radius $R^C_{\rm{out}}$ is an analytic function of $Z_d/Q_p$, thus the squeezing of the barrier is obviously curve dependent on $Z_d/Q_p$ and it increases with $Z_d/Q_p$ in Fig. \ref{fig0}.

\begin{figure}[!]
\centering
\includegraphics[width=8.5cm]{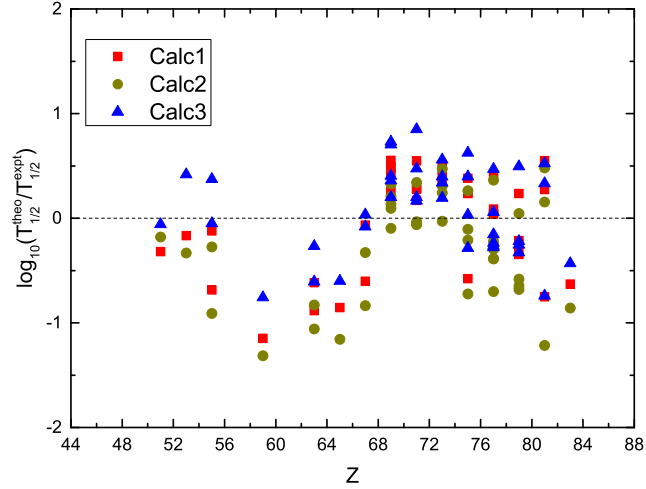}
\caption{\label{fig1}(color online) Decimal logarithm deviations between the experimental data of proton radioactivity half-lives and calculations. The squares, circles and triangles refer to results obtained by our model, Gamow--like model and UDLP, denoted as Calc1, Calc2 and Calc3, respectively.}
\end{figure}

\begin{figure}[!]
\centering
\includegraphics[width=8.5cm]{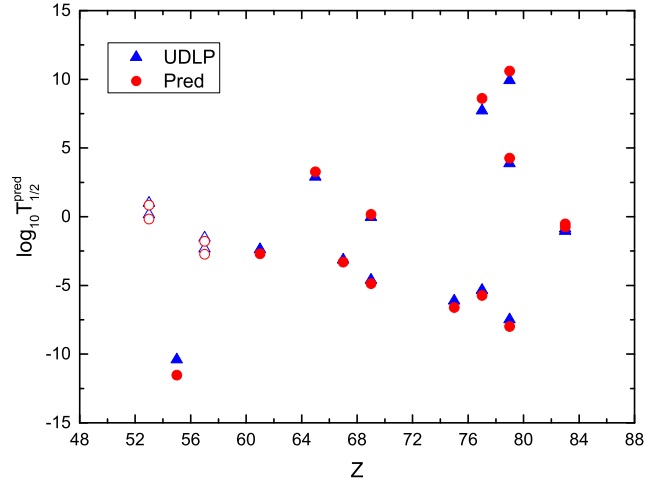}
\caption{\label{fig2}(color online) Comparison of the predicted proton radioactivity half--lives using our model and UDLP. The circles and triangles refer to results obtained by this work and UDLP, denoted as Calc and UDLP, respectively. The hallow circles and triangles denote the values for $^{108}$I and $^{117}$La nuclei with the two cases i.e. $l=$2, 3, respectively.}
\end{figure}

For clearly comparing, we calculate the half--lives of 41 proton emitters by Gamow--like model from Zdeb et al. \cite{Zdeb2016} and universal decay law for proton emission (UDLP) from Qi et al. \cite{PhysRevC.85.011303}. The logarithm of the proton radioactivity half-life obtained by the UDLP can be expressed as
\begin{equation}
\rm{log}_{10}T_{1/2}=A\chi'+B\rho'+C+D\frac{l(l+1)}{\rho'},
\end{equation}
where $A$, $B$, $C$ and $D$ are adjustable parameters, whose values are taken from Ref. \cite{PhysRevC.85.011303}. $\chi'=Z_d\sqrt{A_d/((A_d+1)Q_p)}$, $\rho'=\sqrt{A_dZ_d(A_d^{1/3}+1)/(A_d+1)}$. In addition, we plot the logarithmic differences between $T^{\rm{calci}}_{1/2}$ and  $T^{\rm{expt}}_{1/2}$ versus with proton number of parent nuclei in Fig. {\ref{fig1}} for more detailed comparison. As shown in this figure, these calculated results are well in agree with experimental data, and the results calculated by our model are much better than Zdeb et al. \cite{Zdeb2016}. For systematically comparing, the detailed results including our model, Gamow--like model of Zdeb et al. and UDLP are given in Table {\ref{tab1}}. In this table, the first two columns present the proton emitter and corresponding proton radioactivity energy $Q_p$, respectively. The next two columns denote the transferred minimum orbital angular momentum $l_{\rm{min}}$, the spin and parity transformed from the parent to daughter nuclei, respectively. The last four columns are the half--lives of the experimental data, the calculated results of proton radioactivity half--lives for the present work, Gamow--like model by Zdeb et al. \cite{Zdeb2016} and the UDLP by Qi et al. \cite{PhysRevC.85.011303},  denoted as log$_{10}T^{\rm{calc1}}_{1/2}$, log$_{10}T^{\rm{calc2}}_{1/2}$ and log$_{10}T^{\rm{calc3}}_{1/2}$, respectively. In addition, the standard deviations $\sigma$ from this work, Gamow-like and UDLP are listed in the Table {\ref{tab2a}} for comparison. In this table, the values of $\sigma$ in the first line are calculated from the 41 experimental data used in this work, respectively. While the ones in the second line are taken from Ref. \cite{Zdeb2016} and \cite{PhysRevC.85.011303}. The results show that our improvement in Gamow--like model proposed by Zdeb et al. is obvious, and the calculated half--lives of proton radioactivity are reliable.

\begin{table}[!]
\centering
\caption{\label{tab1}Comparison between the experimental and calculated half--lives of proton radioactivity for $51\leq Z \leq 83$ nuclei. The calculations displayed as log$_{10}T^{\rm{calc1}}_{1/2}$, log$_{10}T^{\rm{calc2}}_{1/2}$ and log$_{10}T^{\rm{calc3}}_{1/2}$ are obtained by our model, Gamow--like model and UDLP, respectively. The symbol $m$ and $n$ denotes the isomeric  state. $'()'$ means uncertain spin and / or parity. $'\#'$ means values estimated from trends in neighboring nuclides with the same $Z$ and $N$ parities.}
\footnotesize
\begin{tabular}{@{}llllllll}
\br
{Nucleus}& $Q_{p}$  & ${J^{\pi}_{p}}\to{J^{\pi}_{d}}$ &$l_{\rm{min}}$&log$_{10}T^{\rm{expt}}_{1/2} $ &log$_{10}{T_{1/2}^{\rm{calc1}}} $ & log$_{10}{T_{1/2}^{\rm{calc2}}} $ & log$_{10}{T_{1/2}^{\rm{calc3}}} $\\  &  (MeV) &   & & (\rm{s})  & (\rm{s})  &  (\rm{s})  &  (\rm{s}) \\
\mr
$^{105}\mathrm{Sb}$ & 0.491 & $5/2^+\to 0^+$ & 2             &2.086&        1.768         & 1.906        &2.027  \\  
$^{109}\mathrm{I}$ & 0.821 & $(3/2^+)\to 0^+$ & 2            &$-$3.897&    $-$4.153     &$-$4.320    &$-$3.569 \\
$^{112}\mathrm{Cs}$ & 0.821 & $1^+\#\to 5/2^+\#$ & 2         &$-$3.310&    $-$3.432     &$-$3.585    &$-$2.936 \\
$^{113}\mathrm{Cs}$ & 0.972 & $(3/2^+)\to 0^+$ & 2           &$-$4.752&    $-$5.438     &$-$5.662    &$-$4.801 \\
$^{121}\mathrm{Pr}$ & 0.891 & $(3/2^+)\to 0^+$ & 2           &$-$1.921&    $-$3.069     &$-$3.236    &$-$2.679 \\
$^{130}\mathrm{Eu}$ & 1.031 & $(1^+)\to(3/2^+,1/2^+)$&2      &$-$3.000&    $-$3.616     &$-$3.830    &$-$3.267 \\
$^{131}\mathrm{Eu}$ & 0.951 & $3/2^+\to 0^+$ & 2             &$-$1.703&    $-$2.586     &$-$2.763    &$-$2.310 \\
$^{135}\mathrm{Tb}$ & 1.181 & $(7/2^-)\to 0^+$ & 3           &$-$2.996&    $-$3.849     &$-$4.152    &$-$3.596 \\
$^{140}\mathrm{Ho}$ & 1.092 & $6^-,0^+,8^+\to (7/2^+)$&3     &$-$2.222&    $-$2.287     &$-$2.549    &$-$2.188 \\
$^{141}\mathrm{Ho^m}$ & 1.251 & $(1/2^+)\to 0^+$ & 0         &$-$5.137&    $-$5.738     &$-$5.972    &$-$5.216 \\
$^{145}\mathrm{Tm}$ & 1.741 & $(11/2^-)\to 0^+$ & 5          &$-$5.499&    $-$5.077     &$-$5.595    &$-$4.796 \\
$^{146}\mathrm{Tm}$ & 0.891 & $(1^+)\to 1/2^+\#$ & 0         &$-$0.810&    $-$0.550     &$-$0.604    &$-$0.405 \\
$^{146}\mathrm{Tm^m}$ & 1.201 & $(5^-)\to 1/2^+\#$ &5        &$-$1.125&    $-$0.629     &$-$1.030    &$-$0.765 \\
$^{147}\mathrm{Tm}$ & 1.059 & $11/2^-\to 0^+$ & 5            &0.573&       1.050         &  0.707        &0.772  \\  
$^{147}\mathrm{Tm^m}$ & 1.120 & $3/2^+\to 0^+$ & 2           &$-$3.444&    $-$2.890     &$-$3.117    &$-$2.713 \\
$^{150}\mathrm{Lu}$ & 1.271 & $(5^-,6^-)\to (1/2^+)$ & 5     &$-$1.201&    $-$0.919     &$-$1.261    &$-$1.002 \\
$^{150}\mathrm{Lu^m}$ & 1.291 & $(1^+,2^+)\to (1/2^+)$ &2    &$-$4.398&    $-$4.226     &$-$4.433    &$-$3.923 \\
$^{151}\mathrm{Lu}$ & 1.243 & $11/2^-\to 0^+$ & 5            &$-$0.916&    $-$0.638     &$-$0.972    &$-$0.747 \\
$^{151}\mathrm{Lu^m}$ & 1.291 & $(3/2^+)\to 0^+$ & 2         &$-$4.783&    $-$4.235     &$-$4.442    &$-$3.933 \\
$^{155}\mathrm{Ta}$ & 1.451 & $(11/2^-)\to 0^+$ & 5          &$-$2.495&    $-$2.139     &$-$2.524    &$-$2.155 \\
$^{156}\mathrm{Ta}$ & 1.021 & $(2^-)\to 7/2^-\#$ & 2         &$-$0.828&    $-$0.438     &$-$0.520    &$-$0.431 \\
$^{156}\mathrm{Ta^m}$ & 1.111 & $(9^+)\to 7/2^-\#$ & 5       &0.924   &     1.437         & 1.167        &1.117  \\  
$^{157}\mathrm{Ta}$ & 0.941 & $1/2^+\to 0^+$ & 0             &$-$0.529&    $-$0.068     &$-$0.057    &0.031  \\  
$^{159}\mathrm{Re^m}$ & 1.816 & $11/2^-\to (0^+)$ & 5         &$-$4.666&    $-$4.425     &$-$4.874    &$-$4.268 \\
$^{160}\mathrm{Re}$ & 1.271 & $(4^-)\to 7/2^-\#$ & 0         &$-$3.164&    $-$3.742     &$-$3.889    &$-$3.448 \\
$^{161}\mathrm{Re}$ & 1.201 & $1/2^+\to 0^+$ & 0             &$-$3.357&    $-$2.974     &$-$3.094    &$-$2.732 \\
$^{161}\mathrm{Re^m}$ & 1.321 & $11/2^-\to 0^+$ & 5          &$-$0.680&    $-$0.443     &$-$0.786    &$-$0.648 \\
$^{164}\mathrm{Ir}$ & 1.844 & $(9^+)\to 7/2^-$ & 5           &$-$3.959&    $-$4.210     &$-$4.661    &$-$4.114 \\
$^{165}\mathrm{Ir^m}$ & 1.721 & $(11/2^-)\to 0^+$ & 5        &$-$3.430&    $-$3.388     &$-$3.819    &$-$3.370 \\
$^{166}\mathrm{Ir}$ & 1.161 & $(2^-)\to (7/2^-)$ & 2         &$-$0.842&    $-$1.094     &$-$1.228    &$-$1.120 \\
$^{166}\mathrm{Ir^m}$ & 1.331 & $(9^+)\to (7/2^-)$ & 5       &$-$0.091&    $-$0.049     &$-$0.387    &$-$0.326 \\
$^{167}\mathrm{Ir}$ & 1.071 & $1/2^+\to 0^+$ & 0             &$-$1.128&    $-$0.716     &$-$0.763    &$-$0.659 \\
$^{167}\mathrm{Ir^m}$ & 1.246 & $11/2^-\to 0^+$ & 5          &0.778   &     0.865         & 0.559        &0.513  \\  
$^{170}\mathrm{Au}$ & 1.471 & $(2^-)\to (7/2^-)$ & 2         &$-$3.487&    $-$3.832     &$-$4.070    &$-$3.708 \\
$^{170}\mathrm{Au^m}$ & 1.751 & $(9^+)\to (7/2^-)$ & 5       &$-$2.975&    $-$3.188     &$-$3.621    &$-$3.228 \\
$^{171}\mathrm{Au}$ & 1.448 & $ (1/2^+)\to 0^+$ & 0          &$-$4.652&    $-$4.415     &$-$4.607    &$-$4.158 \\
$^{171}\mathrm{Au^m}$ & 1.702 & $11/2^-\to 0^+$ & 5          &$-$2.587&    $-$2.842     &$-$3.267    &$-$2.916 \\
$^{176}\mathrm{Tl}$ & 1.261 & $(3^-,4^-,5^-)\to (7/2^-)$ & 0 &$-$2.208&    $-$1.932     &$-$2.053    &$-$1.876 \\
$^{177}\mathrm{Tl}$ & 1.155 & $(1/2^+)\to 0^+$ & 0           &$-$1.178&    $-$0.627     &$-$0.698    &$-$0.654 \\
$^{177}\mathrm{Tl^m}$ & 1.962 & $(11/2^-)\to 0^+$ & 5        &$-$3.459&    $-$4.210     &$-$4.674    &$-$4.200 \\
$^{185}\mathrm{Bi^m}$ & 1.607 & $1/2^+\to 0^+$ & 0           &$-$4.192&    $-$4.822     &$-$5.050    &$-$4.623 \\
\br
\end{tabular}
\end{table}

\begin{table}[!]

\centering
\caption{\label{tab2a}The standard deviations $\sigma$ of this work, Gamow-like and UDLP in different cases}
\footnotesize
\begin{threeparttable}
\begin{tabular}{@{}llll}
\br
Model & This work & Gamow--like  & UDLP  \\
\mr
$\sigma$ & 0.468 & 0.559  &  0.424 \\  
         & & 0.460$\ ^{a}$ &  0.440$\ ^{b}$ \\  
\br
\end{tabular}
\begin{tablenotes}
\item[\it{a}] Taken from \cite{Zdeb2016}.
\end{tablenotes}
\begin{tablenotes}
\item[\it{b}] Taken from \cite{PhysRevC.85.011303}.
\end{tablenotes}
\end{threeparttable}
\end{table}

In the following, we predict the proton radioactivity half--lives of 16 nuclei in region $53\leq Z \leq 83$ within our model, whose proton radioactivity are energetically allowed or observed but  not yet quantified in NUBASE2016 \cite{1674-1137-41-3-030001}. The spin and parity are taken from the NUBASE2016, the proton radioactivity released energies are taken from  the  AME2016 \cite{1674-1137-41-3-030002,1674-1137-41-3-030003}. The results are listed in Table {\ref{tab2}}. In this table, the first four columns are same as Table {\ref{tab1}}, the last three columns are the predicted proton radioactivity half--lives obtained by UDLP and our model and experimental data denoted as $\rm{log}_{10}T_{1/2}^{\rm{UDLP}}$, $\rm{log}_{10}T_{1/2}^{\rm{calc}}$ and $\rm{log}_{10}T_{1/2}^{\rm{expt}}$, respectively. As for $^{108}$I and $^{117}$La, the values of orbital angular momentum $l$ taken away by emitted proton maybe 3 in Ref. \cite{Page1991} and \cite{Sonzogni}, the predictions of the $l=3$ case are also listed in Table {\ref{tab2}}. Based on the $\sigma=0.468$ of our model for 41 nuclei in the same region with predicted proton emitters, thus the predicted proton radioactivity half--lives are within a factor of 2.94. Moreover, we also compare our predicted results with the UDLP. For a more intuitive comparison, we plot the predicted half--lives calculated by our method and UDLP versus with proton number of parent nuclei in Fig. {\ref{fig2}}. The results show that the predicted results by UDLP and our model are consistent.

\begin{figure*}[!]
	\centering
	\subfigure[]{
		\label{fig3a}
		\includegraphics[width=0.4\textwidth]{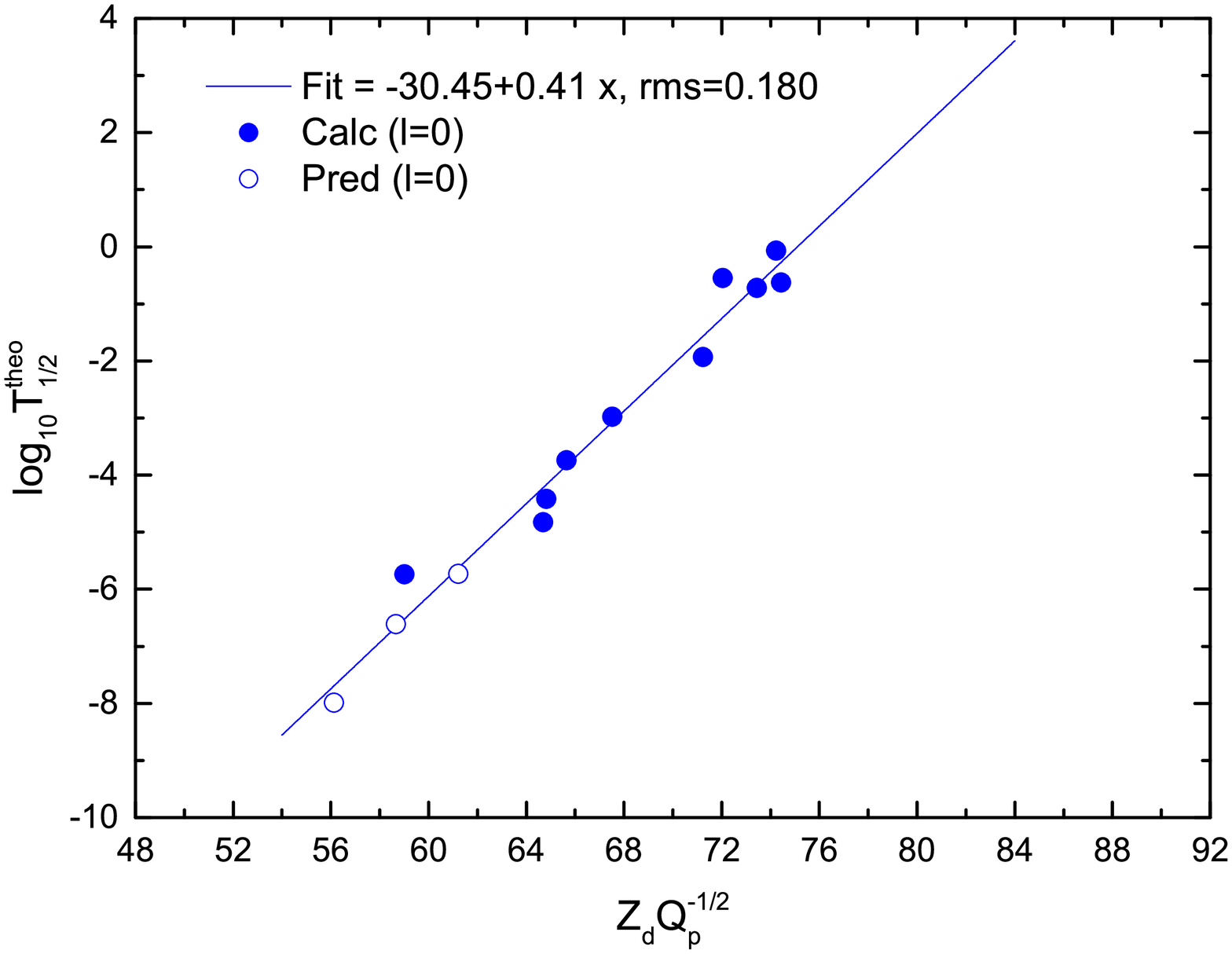}}
	\hspace{1mm}
	\subfigure[]{
		\label{fig3b}
		\includegraphics[width=0.4\textwidth]{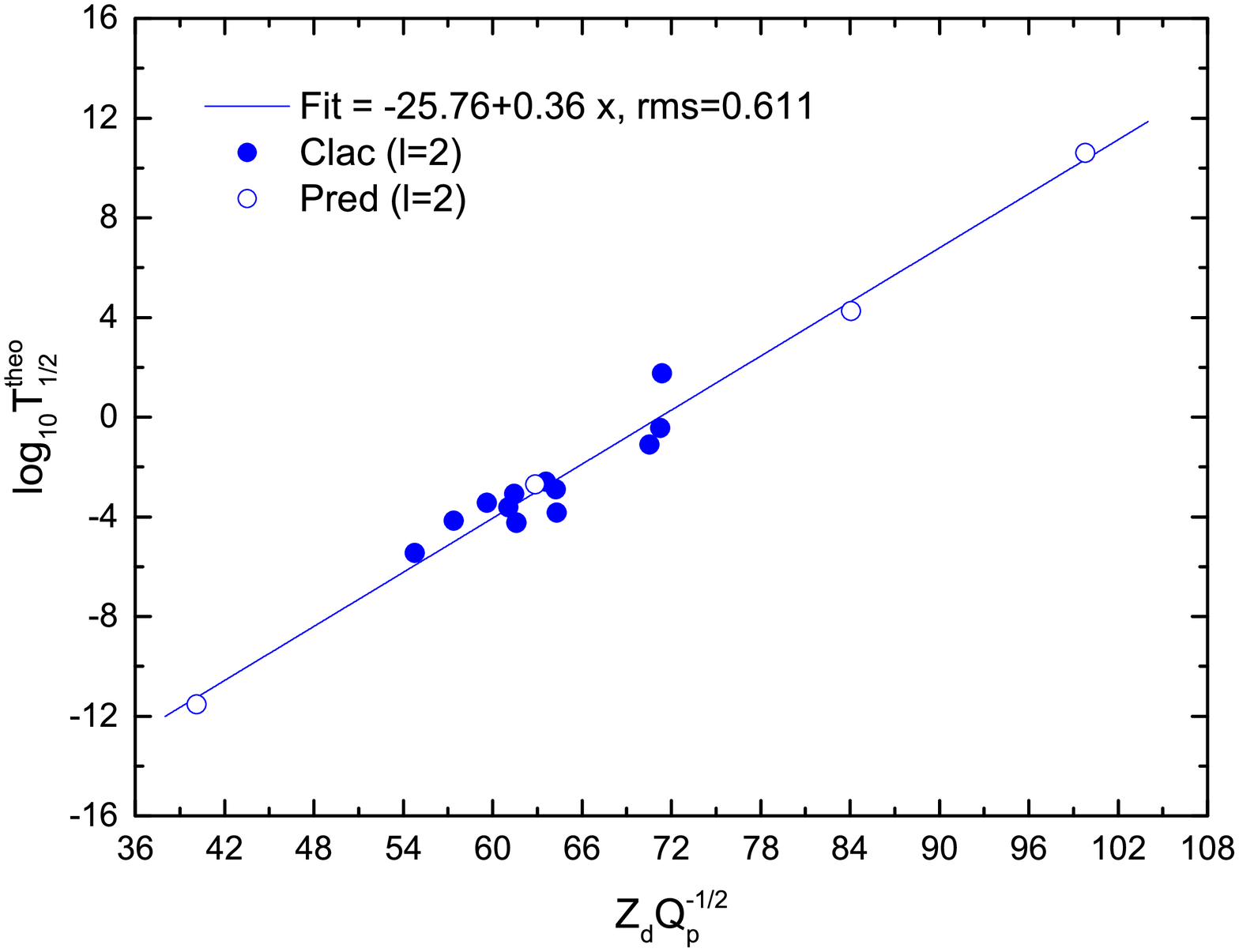}}
	\hspace{1mm}	
		\subfigure[]{
		\label{fig3c}
		\includegraphics[width=0.4\textwidth]{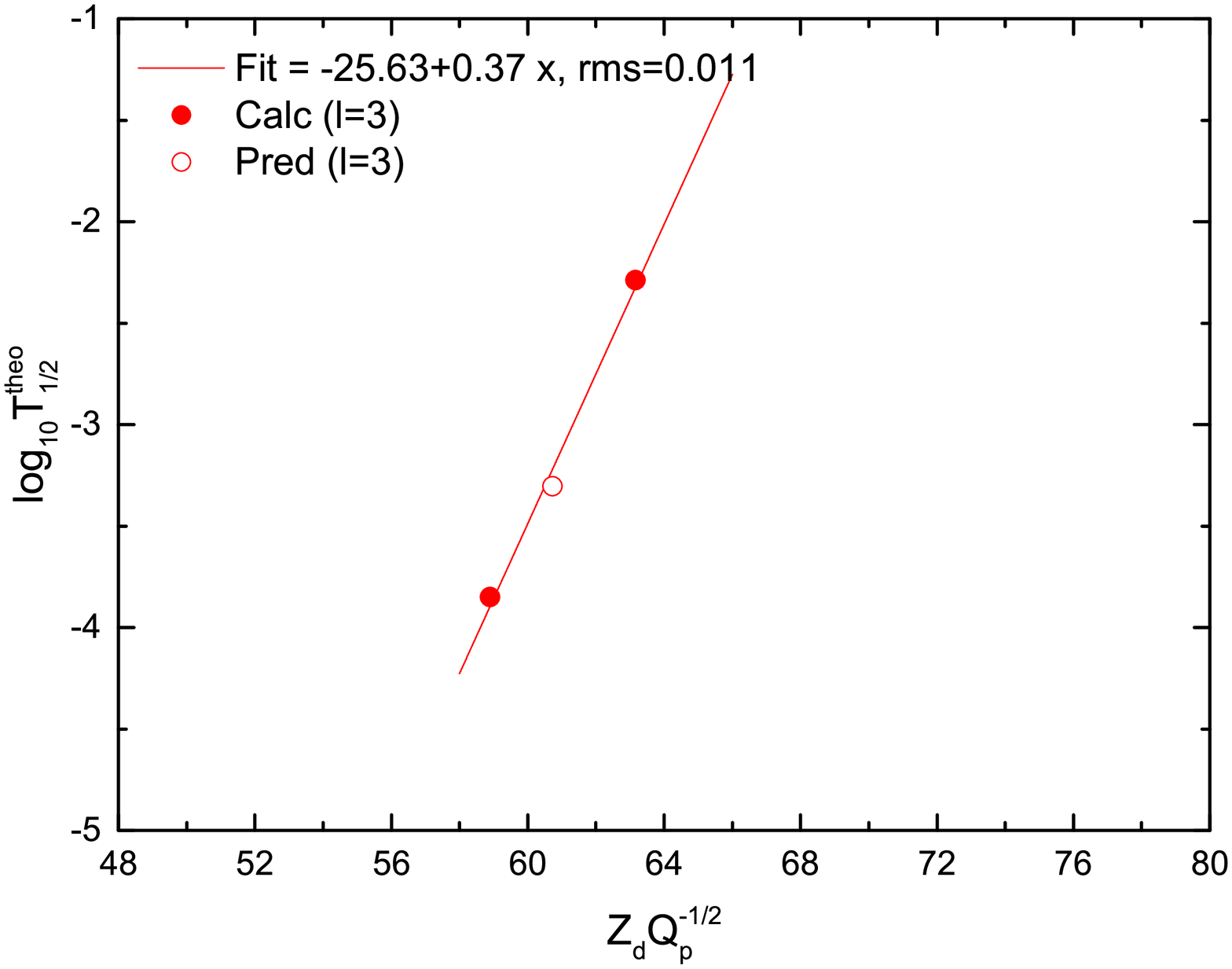}}
	\hspace{1mm}
	\subfigure[]{
		\label{fig3d}
		\includegraphics[width=0.4\textwidth]{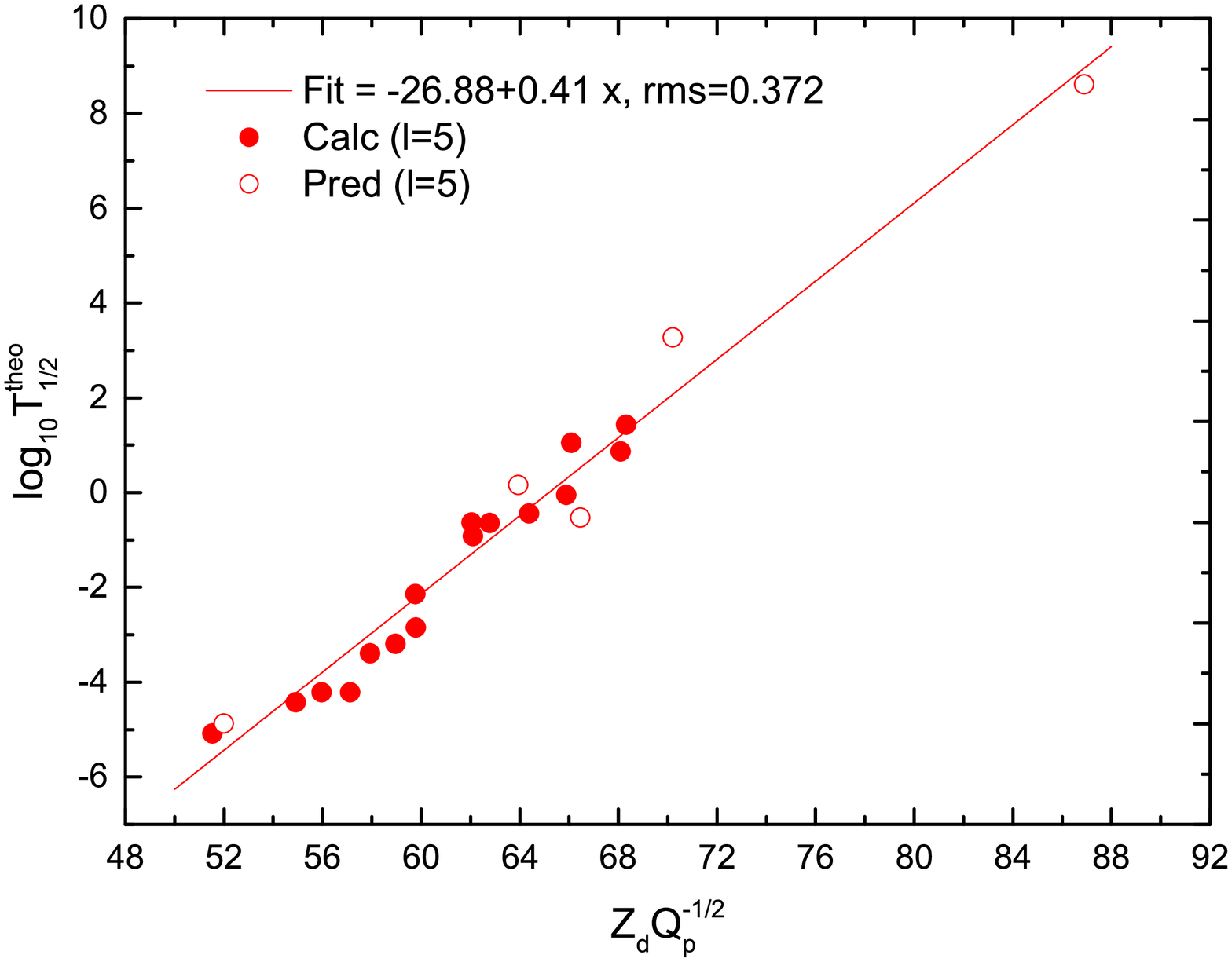}}
	\caption{(color online) The linear relationship between Logarithmic value of half--life $\rm{log}_{10}T_{1/2}^{\rm{theo}}$ and Coulomb parameter $Z_d/\sqrt{Q_p}$. With the orbital angular momentum $l=$0, 2, 3 and 5, the linear relationship are divided into the part of (a), (b), (c) and (d), respectively. The solid and hallow circle denote the calculated half--lives which are taken from Table {\ref{tab1}} and Table {\ref{tab2}}, respectively.}
	\label{fig3}
\end{figure*}

The first striking correlation between the half-lives of radioactive decay processes and the Q--values of the emitted particle was found in $\alpha$ decay by Geiger and Nuttall \cite{doi:10.1080/14786441008637156}. A series of derivation has been developed by studying Geiger--Nuttall law, such as Brown--type empirical formula \cite{PhysRevC.46.811,0954-3899-30-7-011,SILISTEANU20121096,PhysRevC.88.044618,BUDACA201660,Sreeja2018}, the universal decay law \cite{PhysRevC.92.064301,PhysRevLett.103.072501,PhysRevC.80.044326} and its generalization to the proton emission \cite{PhysRevC.85.011303}. Considering the Coulomb parameters $Z_d/\sqrt{Q_p}$ and the effect of orbital angular momentum $l$ which can not be neglected, we plot  the relationships between logarithmic value of our calculated half-lives listed in Table {\ref{tab1}} and {\ref{tab2}} and $Z_d/\sqrt{Q_p}$ in Fig. {\ref{fig3}}. This figure describes four cases i.e. $l=$ 0, 2, 3 and 5 labeled as Fig. {\ref{fig3a}}, Fig. {\ref{fig3b}}, Fig. {\ref{fig3c}} and Fig. {\ref{fig3d}}, respectively. From this figure, we can find that the $\rm{log}_{10}T_{1/2}^{\rm{theo}}$ are linearly dependent on $Z_d/\sqrt{Q_p}$ with the orbital angular momentum $l$ remaining the same. With the change of the orbital angular momentum $l$, the intercept and slope in the figure will be affected and changed in succession. Therefore, the linear relationships between $\rm{log}_{10}T_{1/2}^{\rm{theo}}$ and $Z_d/\sqrt{Q_p}$ can also confirm the above--mentioned statement about the effect of proton radioactivity on orbital angular momentum $l$. Moreover, the well linear relationships indicate that our method is very coincident, and our predicted results are reliable.

\begin{table}[!]
\centering
\caption{\label{tab2}Same as Table {\ref{tab1}}, but for predicted radioactivity half--lives of nuclei in region $53 \leq Z \leq 83$, which proton radioactivity are energetically allowed or observed but not yet quantified in NUBASE2016 \cite{1674-1137-41-3-030001}, within Gamow--like model with a screened electrostatic barrier.}
\footnotesize
\begin{threeparttable}
\begin{tabular}{@{}lllllll}
\br
{Nucleus} & $Q_{p}$  & ${J^{\pi}_{p}}\to{J^{\pi}_{d}}$ &$l_{\rm{min}}$&log$_{10}T^{\rm{UDLP}}_{1/2} $ &log$_{10}{T_{1/2}^{\rm{calc}}} $ &log$_{10}{T_{1/2}^{\rm{expt}}} $\\  &  (MeV) & & & (\rm{s}) & (\rm{s}) & (\rm{s}) \\
\mr
$^{108}\mathrm{I}$ & 0.601 & ($1^+)\#\to 5/2^+\#$ & 2      &  0.164        &$-$0.161 &$>0.556\ ^c$\\ 
                  &        &                    &$3\ ^d$      &  0.996        &0.835   &   \\   
$^{111}\mathrm{Cs}$ & 1.811 & $3/2^+\#\to 0^+$ & 2       &  $-$10.405    &$-$11.521&   \\   
$^{117}\mathrm{La}$ & 0.821 & $(3/2^+)\#\to 0^+\#$&2     &  $-$2.322     &$-$2.728 &$-1.602\ ^e$ \\
                    &       &                     &$3\ ^e$     &  $-$1.530     &$-$1.784 &   \\    
$^{127}\mathrm{Pm}$ & 0.911 & $5/2^+\#\to 0^+$ & 2       &  $-$2.372     &$-$2.694 &  \\   
$^{137}\mathrm{Tb}$ & 0.831 & $11/2^-\#\to 0^+$ & 5      &  2.907        &   3.274 &     \\ 
$^{141}\mathrm{Ho}$ & 1.181 & $(7/2^-)\to 0^+$ & 3       &  $-$3.132     &$-$3.304 &   \\   
$^{144}\mathrm{Tm}$ & 1.711 & $(10^+)\to 9/2^-\#$ & 5    &  $-$4.609     &$-$4.873 & $>-5.638\ ^c$  \\   
$^{146}\mathrm{Tm^n}$ & 1.131 & $(10^+)\to 11/2^-\#$ & 5 &  $-$0.037     &   0.166  &    \\ 
$^{159}\mathrm{Re}$ & 1.591 & $1/2^+\#\to 0^+$ & 0       &  $-$6.121     &$-$6.611  &  \\   
$^{165}\mathrm{Ir}$ & 1.541 & $1/2^+\#\to 0^+$ & 0       &  $-$5.341     &$-$5.728  &  \\   
$^{169}\mathrm{Ir^m}$ & 0.765 & $(11/2^-)\to 0^+$ & 5    &  7.727        &   8.616 &     \\ 
$^{169}\mathrm{Au}$ & 1.931 & $1/2^+\#\to 0^+$ & 0       &  $-$7.483     &$-$7.986 &   \\   
$^{172}\mathrm{Au}$ & 0.861 & $(2^-)\to 7/2^-\#$ & 2     &  3.873        &   4.262 & $>0.146\ ^c$  \\ 
$^{172}\mathrm{Au^m}$ & 0.611 & $(9^+)\to 13/2^+$ & 2    &  9.926        &   10.603 & $>-0.260\ ^c$ \\ 
$^{185}\mathrm{Bi}$ & 1.523 & $9/2^-\#\to 0^+$ & 5       &  $-$0.881     &$-$0.525 &   \\   
$^{185}\mathrm{Bi^m}$ & 1.703 & $13/2^+\#\to 0^+$ & 6    &  $-$1.044     &$-$0.747 &   \\
\br
\end{tabular}
\begin{tablenotes}
\item[\it{c}] Taken from \cite{1674-1137-41-3-030001}.
\end{tablenotes}
\begin{tablenotes}
\item[\it{d}] Taken from \cite{Page1991}.
\end{tablenotes}
\begin{tablenotes}
\item[\it{e}] Taken from \cite{Sonzogni}.
\end{tablenotes}
\end{threeparttable}
\end{table}

\section{Summary}
\label{section 4}
In summary, we systematically study the half--lives of proton radioactivity for $51 \leq Z \leq 83$ nuclei based on Gamow--like model with a screened electrostatic barrier.  The model contains the effective nuclear radius parameter $r_0$ and screening parameter $a$ while considering the screened electrostatic effect of Coulomb potential with the Hulthen potential. By fitting 41 experimental data, the two adjustable parameters are determined as $r_0=1.14$ fm and $a=5.6\times10^{-4}$ fm$^{-1}$, respectively. Our results can better reproduce the experimental data than Zdeb et al.. In this sense, we extend our model to predict proton radioactivity half--lives of 16 nuclei in the same region within a factor of 2.94. In addition, studying on the proton radioactivity half-life by a type of universal decay law indicates that the calculated half--lives are linearly dependent on Coulomb parameter with the same orbital angular momentum.

\section{Acknowledgments}
\label{section 5}

This work is supported in part by National Natural Science Foundation of China (Grant No. 11205083 and No. 11505100), the construct program of the key discipline in Hunan province, the Research Foundation of Education Bureau of Hunan Province, China (Grant No. 15A159 and No. 18A237), the Natural Science Foundation of Hunan Province, China (Grant Nos. 2015JJ3103, 2015JJ2123), the Innovation Group of Nuclear and Particle Physics in USC, the Shandong Province Natural Science Foundation, China (Grant No. ZR2015AQ007).


\section*{References}

\begin{thebibliography}{}


\bibitem{JACKSON1970281}Jackson K, Cardinal C, Evans H, Jelley N and Cerny J 1970 {\it Phys. Lett. B} \textbf{33} 281
\bibitem{CERNY1970284}Cerny J, Esterl J, Gough R and Sextro R 1970 {\it Phys. Lett. B} \textbf{33} 284
\bibitem{Hofmann1982}Hofmann S, Reisdorf W,  M{\"u}nzenberg, F. P. He{\ss}berger G, Schneider J R H and Armbruster P 1982 {\it Z. Phys. A} \textbf{305} 111
\bibitem{Klepper1982}Klepper O, Batsch T, Hofmann S, Kirchner R, Kurcewicz W, Reisdorf W, Roeckl E, Schardt D and Nyman G 1982 {\it Z. Phys. A} \textbf{305} 125
\bibitem{Qian2016}Qian Y B and Ren Z 2016 {\it Eur. Phys. J. A} \textbf{52} 68
\bibitem{1674-1137-41-3-030001}Audi G, Kondev F, Wang M, Huang W and Naimi S 2017 {\it Chin. Phys. C} \textbf{41} 030001
\bibitem{KARNY200852}Karny M \textit{et al} 2008 {\it Phys. Lett. B} \textbf{664} 52
\bibitem{BHATTACHARYA2007263}Bhattacharya M and Gangopadhyay G 2007 {\it Phys. Lett. B} \textbf{651} 263
\bibitem{QIAN-Yi-Bin-72301}Qian Y B, Ren Z Z and Ni D D 2010 {\it Chin. Phys. Lett.}  \textbf{27} 072301
\bibitem{PhysRevC.72.051601}Basu D N, Chowdhury P R and Samanta C 2005 {\it Phys. Rev. C} \textbf{72} 051601
\bibitem{QIAN-Yi-Bin**-112301}Qian Y B, Ren Z Z, Ni D D and Sheng Z Q 2010 {\it Chin. Phys. Lett.} \textbf{27} 112301
\bibitem{PhysRevC.56.1762}\AA{}bergS, Semmes P B and Nazarewicz W 1997 {\it Phys. Rev. C} \textbf{56} 1762
\bibitem{PhysRevC.79.054330}Dong J M, Zhang H F and Royer G 2009 {\it Phys. Rev. C} \textbf{79} 054330
\bibitem{0954-3899-37-8-085107}Zhang H F, Wang Y J, Dong J M, Li J Q and Scheid W 2010 {\it J. Phys. G: Nucl. Part. Phys.} \textbf{37} 085107
\bibitem{yzwang2017}Wang Y Z, Cui J P, Zhang Y L, Zhang S and Gu J Z 2017 {\it Phys. Rev. C} \textbf{95} 014302
\bibitem{Routray2011}Routray T R, Tripathy S K, Dash B B, Behera B and Basu D N 2011 {\it Eur. Phys. J. A} \textbf{47} 92
\bibitem{delion424}Delion D S, Liotta R J and Wyss R 2006 {\it Phys Rep}, \textbf{80} 424
\bibitem{PhysRevC.85.011303}Qi C, Delion D S, Liotta R J and Wyss R 2012 {\it Phys. Rev. C} \textbf{85} 011303
\bibitem{Routray2012}Routray T R, Mishra A, Tripathy S K, Behera B and Basu D N 2012 {\it Eur. Phys. J. A} \textbf{48} 77
\bibitem{FERREIRA2011508}Ferreira L, Maglione E and Ring P 2011 {\it Phys. Lett. B} \textbf{701} 508
\bibitem{PhysRevC.71.014603}Balasubramaniam M and Arunachalam N 2005 {\it Phys. Rev. C} \textbf{71} 014603
\bibitem{1674-1137-34-2-005}Dong J M, Zhang H F, Zuo W and Li J Q 2010 {\it Chin. Phys. C} \textbf{34} 182

\bibitem{PhysRevC.95.044303}Sun X D, Deng J G, Xiang D, Guo P and Li X H 2017 {\it Phys. Rev. C} \textbf{95} 044303
\bibitem{PhysRevC.95.014319}Sun X D , Duan C, Deng J G, Guo P and Li X H 2017 {\it Phys. Rev. C} \textbf{95} 014319
\bibitem{PhysRevC.96.024318}Deng J G, Zhao J C, Xiang D and Li X H 2017 {\it Phys. Rev. C}, \textbf{96} 024318
\bibitem{1674-1137-41-12-124109}Deng J G, Cheng J H, Zheng B and Li X H 2017 {\it Chin. Phys. C} \textbf{41} 124109
\bibitem{1674-1137-42-4-044102}Deng J G, Zhao J C, Chen J L, Wu X J and Li X H 2018 {\it Chin. Phys. C} \textbf{42} 044102
\bibitem{doi:10.1142/S0218301318500052}Soylu A 2018 {\it Int. J Mod. Phys. E}, \textbf{27} 1850005
\bibitem{PhysRevC.96.034619}Santhosh K P and Sukumaran I 2017 {\it Phys. Rev. C} \textbf{96} 034619
\bibitem{PhysRevLett.96.072501}Delion D S, Liotta R J and Wyss R 2006 {\it Phys. Rev. Lett.} \textbf{96} 072501
\bibitem{Budaca2017}Budaca R and Budaca A I 2017 {\it Eur. Phys. J. A} \textbf{53} 160
\bibitem{Zdeb2016}Zdeb A, Warda M, Petrache C M and Pomorski K 2016 {\it Eur. Phys. J. A} \textbf{52} 323
\bibitem{PhysRevC.68.034319}Xu C and Ren Z 2003 {\it Phys. Rev. C} \textbf{68} 034319
\bibitem{PhysRevC.69.024614}Xu C and Ren Z 2004 {\it Phys. Rev. C} \textbf{68} 024614
\bibitem{PhysRevC.78.044310}Ni D, Ren Z, Dong T and Xu C 2008 {\it Phys. Rev. C} \textbf{78} 044310
\bibitem{PhysRevC.81.064309}Dong J, Zuo W, Gu J, Wang Y and Peng B 2010 {\it Phys. Rev. C} \textbf{81} 064309
\bibitem{PhysRevC.62.044610}Myers W D and \ifmmode \acute{S}\else \'{S}\fi{}wia\ifmmode \mbox{\c{}}\else \c{}\fi{}tecki W J 2000 {\it Phys. Rev. C} \textbf{62} 044610
\bibitem{zdeb014029}Zdeb A, Warda M and Pomorski K 2013 {\it Phys. Scr. T} \textbf{154} 014029
\bibitem{zedb024308}Zdeb A, Warda M and Pomorski K 2013 {\it Phys. Rev. C} \textbf{87} 024308
\bibitem{MYERS1974186}Myers W D and \ifmmode \acute{S}\else \'{S}\fi{}wia\ifmmode \mbox{\c{}}\else \c{}\fi{}tecki W J 1974 {\it Ann. Phys.} \textbf{84} 186
\bibitem{MYERS1991292}Myers W D and \ifmmode \acute{S}\else \'{S}\fi{}wia\ifmmode \mbox{\c{}}\else \c{}\fi{}tecki W J 1991 {\it Ann. Phys.} \textbf{211} 292
\bibitem{Hulthen1942}Hulthen L 1942 {\it Ark. Mat. Astron. Fys. A} \textbf{28} 52
\bibitem{Hulthen1957}Hulthen L, Sugawara M, S.Flugge {\it (Editors)} 1957 {\it Handbuchder Physik (Springer)}
\bibitem{Oyewumi2016}Oyewumi K J and Oluwadare O J 2016 {\it Eur. Phys. J. Plus} \textbf{131} 295
\bibitem{doi:10.1063/1.531270} Morehead J J 1995 {\it J. Math. Phys.} \textbf{36} 5431

\bibitem{1674-1137-41-3-030002}Huang W, Audi G, Wang M, Kondev F, Naimi S and Xu X 2017 {\it Chin. Phys. C} \textbf{41} 030002 
\bibitem{1674-1137-41-3-030003}Wang M, Audi G, Kondev F, Huang W, Naimi S and Xu X 2017 {\it Chin. Phys. C} \textbf{41} 030003
\bibitem{Page1991}Page R D, Woods P J {\it et al} 1991 {\it Z. Phys. A} \textbf{338} 295
\bibitem{Sonzogni}Sonzogni A A 2002 {\it Nucl. Data Sheets} \textbf{95} 1
\bibitem{doi:10.1080/14786441008637156}Geiger H and Nuttall J M 1911 {\it Philos. Mag.} \textbf{22} 613
\bibitem{PhysRevC.46.811}Brown B A 1992 {\it Phys. Rev. C} \textbf{46} 811
\bibitem{0954-3899-30-7-011}Das S and Gangopadhyay G 2004 {\it J. Phys. G: Nucl. Part. Phys.} \textbf{30} 957
\bibitem{SILISTEANU20121096}Sili\ifmmode \mbox{\c{s}}\else \c{s}\fi{}teanu I and Budaca A I 2012 {\it At. Data and Nucl. Data Tables} \textbf{98} 1096
\bibitem{PhysRevC.88.044618}Budaca A I and Sili\ifmmode \mbox{\c{s}}\else \c{s}\fi{}teanu I 2013 {\it Phys. Rev. C} \textbf{88} 044618
\bibitem{Sreeja2018}Sreeja I and Balasubramaniam M 2018 {\it Eur. Phys. J. A} \textbf{54} 106
\bibitem{BUDACA201660}Budaca A I, Budaca R and Sili\ifmmode \mbox{\c{s}}\else \c{s}\fi{}teanu I 2016 {\it Nucl. Phys. A} \textbf{951} 60
\bibitem{PhysRevC.92.064301}Wang Y Z, Wang S J, Hou Z Y and Gu J Z 2015 {\it Phys. Rev. C} \textbf{92} 064301
\bibitem{PhysRevLett.103.072501}Qi C, Xu F R, Liotta R J and Wyss R 2009 {\it Phys. Rev. Lett.} \textbf{103} 072501
\bibitem{PhysRevC.80.044326}Qi C, Xu F R, Liotta R J, Wyss R, Zhang M Y, Asawatangtrakuldee C and Hu D 2009 {\it Phys. Rev. C}, \textbf{80} 044326 


\end{thebibliography}
\end{document}